# Music Tempo Estimation via Neural Networks – A Comparative Analysis

**Mila Soares de Oliveira de Souza***, **Pedro Nuno de Souza Moura***, **Jean-Pierre Briot****

* Escola de Informática Aplicada – Universidade Federal do Estado do Rio de Janeiro (UNIRIO)
22290-255 Rio de Janeiro, RJ, Brazil

** Sorbonne Université, CNRS, LIP6, F–75005 Paris, France

milasoaresdeoliveira@gmail.com, pedro.moura@uniriotec.br, Jean-Pierre.Briot@lip6.fr

**Abstract**

This paper presents a comparative analysis on two artificial neural networks (with different architectures) for the task of tempo estimation. For this purpose, it also proposes the modeling, training and evaluation of a B-RNN (Bidirectional Recurrent Neural Network) model capable of estimating tempo in bpm (beats per minutes) of musical pieces, without using external auxiliary modules. An extensive database (12,550 pieces in total) was curated to conduct a quantitative and qualitative analysis over the experiment. Percussion-only tracks were also included in the dataset. The performance of the B-RNN is compared to that of state-of-the-art models. For further comparison, a state-of-the-art CNN was also retrained with the same datasets used for the B-RNN training. Evaluation results for each model and datasets are presented and discussed, as well as observations and ideas for future research. Tempo estimation was more accurate for the percussion-only dataset, suggesting that the estimation can be more accurate for percussion-only tracks, although further experiments (with more of such datasets) should be made to gather stronger evidence.

**1. Introduction**

Music Information Retrieval (MIR) is the field of study that aims to extract, analyze and provide information from a song [1]. Basic methodologies for Music Information Retrieval can include audio signal processing, musical perception, among others [2]. Some examples of MIR research involve tempo estimation [3] — which is the main research theme of this paper —, identifying musical styles [4] and comparison of similarity between two songs [5]. Estimating the tempo of a song is considered one of the most fundamental tasks of MIR [6].

The tempo of a musical piece corresponds to the number of beats that would be counted in one minute (beats per minute, or bpm) [7]. In general terms, it can be understood as the speed at which humans often tap their fingers or feet as they listen to music [6]. When listening to a song, it is often generalized that music sounding "fast" to a listener should have a longer tempo than songs that seem comparatively slower.

Rhythm is one of the most important characteristics of music [7], and it can be said that it is one of the factors that makes certain musical genres easily recognizable (e.g.: waltz in 3/4 and rock in 4/4-time signatures). Some of them more often work with faster tempo – like heavy metal – while others often work with comparatively shorter tempo – like reggae. Identifying such tempo could have interesting applications in other contexts, such as, for example, a recommendation system for similar songs, which would suggest tracks considering similar tempi and/or other characteristics.

As such, it could be interesting for research methods to estimate the tempo of a musical piece using modern technologies with good accuracy, as well as verifying if there is any model in such technology that stands out in performance. Still considering the importance of tempo, it would be also interesting to test whether the tempo estimation is more accurate when estimating over tracks containing percussion only, which is often one of the most responsible instrument(s) for determining tempo.

The tempo estimation in music using computational methods has been studied for some decades. As an example, the work of Schreirer [8], from 1998, was one of the first to process the audio signal in a continuous way (previously, the usual methods worked over discrete temporal events) [9].

As for tempo estimation using artificial neural networks (ANNs), the first evidences were found at late 2000s-early 2010s. Some of these works were those of Böck [6], circa 2010, and that of Gkiokas, Katsouros and Carayannis [10] in 2012. Böck uses a bidirectional Long-Short Term Memory (LSTM) [11] in order to map beat activations, after which the results are further processed with comb filters. The work of [10] also uses a convolutional neural network to derive a beat activation function, then identify the tempo in later steps.

A neural network model considered state of the art (for tempo estimation) was proposed by Böck and Schedl in 2011 [12] and further refined in 2015 [9]. The model is a bidirectional LSTM, aided by external modules which process detected beat activations. The first published model capable of estimating tempo without the help of external modules that presented state of the art performance was the work of Schreiber and Müller [13] published in 2018, consisting of convolutional and dense layers. This model was an important starting point for the development of this work. The model on [13] presented results comparable to the state of the art. Böck and Davies, in 2020 [14], proposed a system which uses a CNN with convolutional layers and TCN layers (temporal convolutional networks), then processes the results with deep Bayesian networks, also achieving state-of-the-art results.

We propose the use of datasets consisting of percussion instruments only (such as drums) to the estimation of tempo via neural networks research. One of the reasons this is yet to be more explored might be due to the lack of extensive availability of such databases to the MIR community. However, we found there are percussion-only databases being constructed, which already provide a reasonably good number of tracks and annotations to work with. As such, we decided to explore this possibility.

This paper also aims to propose, train and evaluate a new bidirectional recurrent neural network model (B-RNN) that is capable of estimating the tempo in bpm of a musical piece without the need for help of other additional modules (external to the model itself). We aim to evaluate the performance over a collection of datasets, then compare and discuss the results considering other architectures such as convolutional neural networks.

## 2. Dataset

For the experiment, we built a large dataset containing complete musical pieces (or samples). These pieces might contain either multiple instruments or a single instrument, such one drum line (which is a percussion instrument) without other instruments.

| Name | Qt. of samples | Dura-tion (s) | Extension | Genres |
|---|---|---|---|---|
| ACM [15] | 1.410 | 30 | .wav | Pop, rock |
| Extended Ballroom [16] | 3.826 | 30 | .mp3 | Salsa, foxtrot, samba etc. |
| GiantSteps Tempo [17] | 664 | 120 | .mp3 | EDM |
| GiantSteps MTG[a] | 1.158 | 120 | .mp3 | EDM |
| Groove[b] | 443 | 10-60+ | .wav | Reggae, pop, rock, jazz etc. |
| GTZan[18] | 999 | 30 | .wav | Pop, rock etc. |
| Hainsworth [5] | 222 | 40-60+ | .wav | Folk, jazz etc. |
| LMD [19] | 3.611 | 30 | .mp3 | Pop, rock, classical etc. |
| SMC [20] | 217 | 40 | .wav | Classical, romantic, acoustic etc. |

[a.] Available on: <https://github.com/GiantSteps/giantsteps-tempo-dataset>. Accessed: 30 April 2021.
[b.] Available on: <https://ai.google/research/teams/brain/magenta/>. Accessed: 30 April 2021.

**Table 1: Selected datasets and their features.**

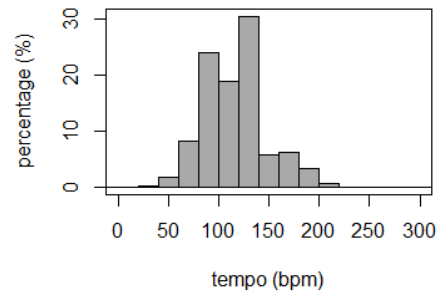

**Figure 1: Tempo distribution of the dataset samples.**

The dataset for this experiment contains 12,550 parts/samples. It is a collection of smaller datasets that were selected for their variety of pieces (diversity of musical genres, also avoiding repetition of data). The free availability for use in academic experiments — as well as availability of public annotations (of tempo in bpm) — in their respective papers/source were also important factors.

Datasets containing percussion-only tracks were also considered necessary for this experiment. This is due to the fact that percussion instruments, such as drums, are often the most used instrument to guide the rhythm of a song (particularly in pop and rock); thus, it of our interest to verify whether the tempo estimation could be more accurate if the network tries to estimate the tempo only over the percussion. Furthermore, to the best of our knowledge there are no works on tempo estimation (with neural networks) that presented experiments with drum-only tracks (nor the Groove dataset). Should the network performance over drums-only pieces eventually be considerably more accurate, it could be interesting to try extracting the drum lines from a multi-instrument song for future experiments.

Table 1 lists the names, number of pieces and presents a brief description of each database, showing also the musical genres of the pieces that make up each one. After the name of each dataset, there is also a reference to their original publication, which grants the access to the musical pieces and the available annotations (exceptions: the annotations for GiantSteps MTG and LMD were obtained from [13] instead of their original sources).

The Groove dataset is the only that contains purely percussion pieces (in this case, drums), totaling 443 samples. All the others present mostly multi-instrument tracks, some of them containing vocal lines. This can be considered an obstacle for the goal of determining whether there is greater accuracy in estimating the tempo of percussion-only pieces, since the amount of data is a crucial factor for the performance of a neural network [21]. Although the original Groove dataset on the source page is more extensive, most of the samples which are available in the Grove dataset consist of information that is not relevant to this experiment, such as drum fills. The authors filtered the database and annotations so that only tracks with proper instrumentation (and available .wav file) remained.

It is interesting to highlight a few points on the selected datasets: Extended Ballroom contains music from 13 different genres, as well as time signatures other than 4/4 (which is the most common in pop songs), such as 3/4. SMC, according to the authors [20], was created with the intention of being particularly challenging for rhythm analysis systems. It includes samples of several styles such as classical music, blues, soundtracks, etc. Groove also features popular drum lines in a variety of genres such as punk, rock, jazz, and gospel, also presenting a variety of time signatures including 3/4 and 6/8. It is possible to consider that Groove does not suffer from style bias, as the tracks were performed by several drummers. The diversity of styles and metrics is an enriching factor for the database.

## 3. Methodology

Tempo estimation of a musical piece (which is originally an audio file) can be approached as both a regression problem and a classification problem. The classification approach can consider the tempo in bpm of a song as a class; a class is an integer number which is between the lowest and the highest bpm values selected. This method could be interesting due to the possibility of working later with other tempo probabilities, such as finding a second or third tempo considered more suitable [13]. This information could also be relevant when analyzing a musical piece structured with multiple tempi (although this paper proposes to work with the scope of a single tempo only). Furthermore, decimal values can be deemed irrelevant for tempo estimation. For these reasons, the authors chose to work with the classification problem approach instead of regression.

It is worth noting that music does not necessarily consist of a single tempo in bpm. There are pieces that present bpm changes over time: for example, in classical music, it is common to have different tempi in distinct parts of the piece's structure. Therefore, it is to be expected that the estimation of a single tempo will not be accurate for such pieces. It is expected, for this experiment, that the neural network will be able to estimate with some precision the tempo of music that present a single tempo for most of their duration.

After defining that the experiment will be conducted on neural networks, the first step was the dataset selection (introduced in Section 2). Then the following steps consisted of defining the input representation for the neural network, defining the neural network model, network training and evaluation; these first three following steps are presented and discussed in this Section, while the evaluation of models and results are discussed in Section 4.

In short, we obtain a Mel spectrogram of the music piece chosen as input, scale it on the time axis, crop it into several windows. These windows are the input for the neural network. The model will then process these windows, estimating the probabilities of the tempo classes for the windows; Finally, a single value of tempo in bpm will be chosen as output after considering all the windows probabilities. These processes and definitions will be further explained in this Section.

### 3.1 Input representation

For this experiment, the Mel spectrogram (which will be defined in the next paragraphs) was selected as the input representation for the neural network. The input treatment process that will be described is replicated from [13].

Although the most basic form of signal representation is the waveform, which represents the magnitude (y-axis) over time, one of its downsides is being quite heavier than other representations for a system to process. Therefore, it was on the interest of the authors using Fourier transforms to obtain a spectrogram, which works in the context of frequency.

The spectrogram provides a two-dimensional visual representation of an audio signal, in which the horizontal axis represents time and the vertical axis represents frequency [22], while color schemes can be used to represent intensity (third axis). Figure 2 is an example of a spectrogram.

The Mel spectrogram is a spectrogram that uses the Mel scale which represents frequencies akin to human perception [23]. The Mel scale is widely used in MIR research. Since music is created by human beings, rhythm and time are created and perceived by human beings as well. Therefore, the Mel scale is useful to carry out experiments that aim to simulate hearing in a closer way to that of its creators.

It is possible to process the input further for network training efficiency by scaling the spectrogram. The entire musical piece spectrogram can be scaled on the time axis (keeping the frequency axis untouched). For this experiment, the spectrogram is scaled along the time axis using a random factor $c \in \{0.8, 0.84, ..., 1.16, 1.2\}$ (the tempo ground truth label is also scaled accordingly to factor c). This also generates diversity of inputs for further training of the network.

Instead of providing an entire spectrogram for the neural network to process, it is possible to crop several windows of this same spectrogram, which was the selected input method in this work. This mode has not only an advantage from a size efficiency point of view (input gets much smaller), but also from the number of examples for training. Cropping the spectrogram in windows, it is possible to generate several different training examples for the network from what was, originally, only one input.

The duration of 10 seconds can be considered enough to get a good sense of the tempo of a piece. Therefore, it is enough to choose a number of frames that is sufficiently close to such time interval. The value of 256 frames (approximately 11.9 seconds) for the spectrogram input, used by [25], was considered adequate.

Thus, the input to the neural network is a honey spectrogram with dimensions $F_T \times T_T = 40 \times 256$ ($F_T$ is the number of *frequency bins* and $T_T$ is the number of time frames). $F_T$ covers the frequency range between 20 and 5,000 Hz.

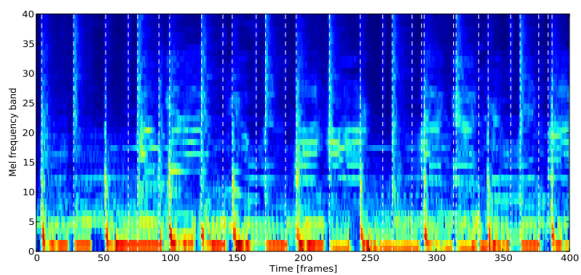

**Figure 2: A Mel spectrogram (extracted from [6]).**

There are also some interesting variations of the Mel spectrogram, such as the log Mel (which uses a logarithmic scale for intensity, that is, decibels). It may be interesting, for a future experiment (not in the scope of this paper), to use the log Mel spectrogram pursuing improvements in tempo estimation.

### 3.2 Model

For this experiment, a new model of bidirectional recurrent neural networks is proposed. Bidirectional recurrent neural networks (B-RNN), introduced by Schuster and Paliwal [25], are recognized as a good model for tasks such as speech recognition [26], also proving to be valuable for tempo estimation. Böck and Schedl [12] propose a B-LSTM (bidirectional Long Short-Term Memory) (which is a type of B-RNN [26]) for tempo estimation; their model can provide state of the art accuracy [13], although they use external modules to process beat activation functions, which is not the goal of this paper.

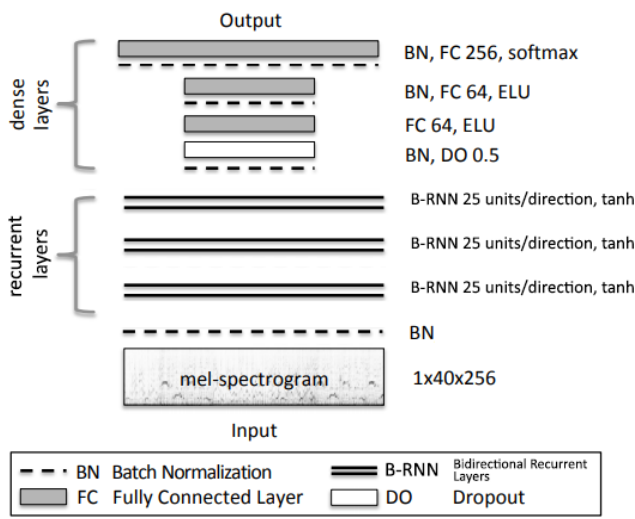

**Figure 3: The B-RNN model overview (modified from [13]).**

The use of bidirectionality makes sense, since, when analyzing a song, both the previous context and the future context of a moment in a song can be used to help determine the tempo. Considering the aforementioned factors, the bidirectional recurrent neural network was chosen as a suitable type of model to be worked with in this paper.

Since the goal is to estimate tempo without the need of auxiliary modules external to the neural network itself, and the authors could not find any B-RNN models yet published that perform this task, a new model was proposed. The overview on the model is represented on Figure 3.

We constructed 3 B-RNN layers with 25 simple recurrent units for each direction to process the input. This corresponds to 6 layers and 150 units. The number of units was inspired on Böck and Schedl [12] B-LSTM. Since the B-RNN is simpler than the B-LSTM, it might be interesting to try more layers or units in the future (which goes beyond the scope of this paper). The recurrent layers have tanh (hyperbolic tangent) as the activation function. The input is normalized before it is sent to the recurrent layers.

The goal for recurrent layers is to identify onsets, analyzing the Mel spectrogram frequencies, and identifying its temporal dependencies. Onset is considered the exact moment of the beginning of a beat; detecting them is needed to find its periodicities. With the bidirectional recurrency, we expect to detect sufficiently long temporal dependencies.

Since we do not want to use external modules for signal processing (such as comb filters), there is the need for the model to be able to process the output from the recurrent layers. Schreiber and Müller [13] proposed a CNN that use dense layers for such feature. Since their model also presented state-of-the-art results, we decided to replicate the dense layers from [13]. The replicated dense layers can also be seen in Figure 3.

Thus, the next step is the processing by these dense layers, whose purpose is to classify the features detected by the recurrent layers. First, the output of the recurrent layers undergoes an average pooling of $(5 \times 1)$. After a batch normalization as the next step, a Dropout layer (p = 0.5) is added to avoid overfitting, succeeded by a dense layer. A batch normalization followed by dense layers will then finish processing the data. The first two dense layers use ELU (exponential linear unit) as the activation function, while the last one uses softmax.

Since the network deals with a multiclass classification problem, we chose categorical cross entropy for the error function. The chosen optimizer is SGD (stochastic gradient descent) with a clipping value of 5 to prevent gradient explosion. The value for learning rate is 0.001; for momentum it is 0.9. The bidirectional recurrent neural network has a total of 6.583.772 trainable parameters.

### 3.3 Training

For our experiment, it is necessary to select a part of the dataset that will be used to train the neural network. Thus, we must divide it in two distinct subsets: one for the training step, and another for the evaluation step later.

In order to avoid biased results, it was decided that datasets in the training subset cannot be in the test subset

(and vice versa), with the exception of Groove (because it is the only one that contains drum-only tracks). It is desirable that the training subset contains a good variety of genres.

Based on these criteria, we selected for training the models the following subsets: Extended Ballroom, GiantSteps MTG, Hainsworth, LMD and part of Groove (90%), totaling 9,215 samples.

Among these datasets selected for training, it is necessary to split them into two more subsets, one which will be used for the network training itself and the other one which should be used to measure the validation loss. The 80% rate was set for the model training, with then 20% for validation. The tracks for each subset were selected randomly.

The only exception is Groove, as mentioned: since it does not contain an overwhelming number of samples (443 pieces only), also being the only dataset with percussion lines, we must consider a subset for evaluation. Groove was split following the proportions: 80% training, 10% validation and 10% evaluation. We made an effort to ensure that tracks used in evaluation were as different as possible from those in training and validation.

As discussed in Section 3.1, the Mel spectrogram of the entire signal is compressed (or expanded, if the duration of the entire audio is less than 11.9 s) and then cropped into smaller excerpts of approximately 11.9 seconds. This process therefore increases the number of inputs that are provided to the network during training.

The training was carried out under early stopping (when there is no validation loss in the last 100 Epochs). The code used to run the training stage was originally made available in [24], with some minor modifications made by us to allow the use of audio files not supported by the original repositories.

The B-RNN training lasted a total of 410 Epochs (in 25 hours), ending due to early stopping. We trained the model only once, with no additional retraining.

### 3.4. Estimating tempo

As stated in Section 3.1, the whole musical piece spectrogram is cropped before being fed to the neural network. Thus, the network will estimate the tempo of each excerpt picking the class with the greatest activation.

To estimate the tempo for the whole track, multiple output activations are calculated using a sliding window with half-overlap (hop-size of 128 frames, approximately 5.96 s). The activations are averaged class-wise and the tempo class with the greatest activation is picked as well. This methodology is replicated from [13].

### 4. Evaluation and results

Following the training stage criteria, the selected datasets for the neural network evaluation step were ACM, GiantSteps Tempo, GTzan, SMC and part of Groove (10%), totaling 3,335 different pieces.

To evaluate the performance, we first need to define what is accuracy in tempo estimation. The most objective concept is to estimate exactly the same value (an integer number such as 80 bpm). However, the difference between songs with close tempo values (e.g. 83 bpm and 85 bpm) can be practically imperceptible to human ears, so a small margin of error must also be taken into account, since it does not compromise the understanding of the music rhythm. As such, more than one level of accuracy for the measurement (secondary accuracies) are usually introduced in tempo estimation studies [27].

It is also common for machines to end up interpreting the value of tempo as 2-3 times greater (or smaller) than the time perceived by humans, a phenomenon known as tempo octave error [28], which is also a research subject in MIR. In fact, human beings themselves often disagree as to the value of tempo perceived by each one [27]. One person might, for example, interpret the tempo of a song as 60 bpm, while another one interprets it as 120 bpm. Therefore, even though the best performance would be as close as possible to the true tempo value, it is also of interest to simulate human hearing and its inaccuracies, without discarding multiples of the original tempo.

We defined some support variables to estimate the accuracy of tempo detection (in bpm) by the neural network. 3 types of accuracy were chosen: Accuracy0, Accuracy1 and Accuracy2. It can be said that Accuracy0 and Accuracy1 are the ones that represent the best performance as they are closer to the original absolute value; Accuracy2 represents a performance that can be considered good enough (and could be refined in future work, not in the scope of this paper). These accuracies refer to the tempi values estimated by the neural networks for one or more musical pieces.

Accuracy0 is the accuracy rate for tempo values (rounded to the nearest whole number) identical to the ground truth label; Accuracy1 accepts estimations with a deviation of ±4% from the ground truth label; and Accuracy2 accepts values two or three times greater (or smaller) than the ground truth label, plus a ±4% margin of error.

These variables were inspired by studies such as [27] and, mainly, by [13], from which the margin of error percentage (±4%) was replicated. The criteria for each accuracy were chosen not only for they are suitable for an accuracy analysis, but also to facilitate a comparison of the results with a state-of-the-art model.

| Dataset | B-RNN | Sch1 [24] | Sch2 |
|---|---|---|---|
| *ACM [15]* | 33.0 | **40.6** | 39.3 |
| *Groove* | 58.1 | 37.2 | **60.6** |
| *GiantSteps Tempo [17]* | 15.7 | 27.6 | **27.7** |
| *GTzan [18]* | 25.2 | **36.9** | 30.5 |
| *SMC [20]* | 6.0 | **12.4** | 11.1 |

**Table 2: Accuracy0 results for the compared models.**

| Dataset | B-RNN | Sch1 [24] | Sch2 |
|---|---|---|---|
| *ACM [15]* | 72.1 | **79.5** | 73.8 |
| *Groove* | **76.7** | 62.8 | 72.1 |
| *GiantSteps Tempo [17]* | 69.3 | 64.6 | **83.0** |
| *GTZan[18]* | 62.0 | **69.4** | 64.6 |
| *SMC[20]* | 18.4 | **33.6** | 27.2 |

**Table 3: Accuracy1 results for the compared models.**

| Dataset | B-RNN | Sch1 [24] | Sch2 |
|---|---|---|---|
| *ACM [15]* | 93.1 | **97.4** | 96.5 |
| *Groove* | **95.4** | 86.0 | 93.0 |
| *GiantSteps Tempo [17]* | 86.3 | 83.1 | **92.5** |
| *GTZan[18]* | 85.2 | **92.6** | 91.0 |
| *SMC[20]* | 30.4 | **50.2** | 40.5 |

**Table 4: Accuracy2 results for the compared models.**

The results for the B-RNN are compared to those presented by Schreiber and Müller's [13] CNN (convolutional neural network), which we will refer to as Sch1. Since Sch1 presented state-of-the-art results and it was the first not to recur to external modules in tempo estimation, we will be including their results in the analysis; for a comparison of neural networks-only vs. neural networks with external modules, it is recommended to read their original paper [13].

For comparison purposes, the CNN model from [24] was reproduced and trained from scratch with the same datasets and hyperparameters used for the B-RNN of this paper. We will refer to this retrained model as Sch2. This retraining was done as an effort to avoid a possibility of bias in estimating percussion tracks (since Sch1 was not trained with this particular type of dataset).

Tables 2, 3 and 4 each show the percentage of correct estimations for Accuracy0, Accuracy1 and Accuracy2, respectively. Best results for each dataset are in bold.

The B-RNN presented better results than Sch1 [13] in all accuracies for the Groove dataset, and better results in Accuracy1 and Accuracy2 for the GiantSteps Tempo dataset. As previously stated, Sch1 was not trained with the Groove dataset.

Nevertheless, B-RNN's performance on the Groove dataset (which contains only drum-line audios) was superior to that of CNNs: of the three types of accuracy established, B-RNN had the best result in two. This can suggest that B-RNNs might detect temporal dependencies better than CNNs on percussion-only tracks. Still, more research should be done on this matter for stronger evidence. Furthermore, Groove was the dataset that received the best accuracies for the models in this experiment.

The Groove dataset had relatively high Accuracies 0, 1 and 2 values for the B-RNN and Sch2 models, which were trained with percussion pieces. In fact, it was the dataset that presented the highest Accuracy0 values, and the only dataset in which B-RNN surpassed Sch1 and even Sch2 (which was also trained with Groove) in accuracy performance, as mentioned already in this section. As the only percussion dataset, containing 443 tracks — that is, only 3.53% of the total pieces —, we had expected the networks would not perform very well compared to the multi-instrumented datasets (which are 12,107 tracks, 96.47% of the total). Obviously, the possibility of bias should be considered since part of the dataset was used for training and the other part for the assessment itself.

Some of the B-RNN's inaccuracies on Groove were a 145-bpm estimation provided for a 290 bpm song. 290 bpm is a very unusual tempo, not only for the training dataset distribution but for music as a whole, and that might have contributed to the choice of 145 instead of 290. Another estimation error was the attribution of 84 bpm to a track that has 60 bpm, but which has an unusual metric of 6/8, which may have contributed to the error; yet, for another song in 6/8 metric the detection was accurate enough (a 70-bpm track was rated 140 bpm, double the real value).

The highest Accuracy0 value, after Groove, is for the ACM dataset. It makes sense since most of its content is pop and pop-rock songs, which are usually well-defined pieces with few tempo changes. Accuracy2 had the highest values for all Accuracies, which is expected, since it is the least "rigorous" secondary accuracy. Still, since it exceeded 85% accuracy for all datasets (except for the SMC, which was designed to be challenging), it is possible to say that the B-RNN model presents satisfactory results, consistent to human hearing.

It is noticeable that, despite the two CNNs (Sch1 and Sch2) having exactly the same architecture, their performances presented different accuracies for the same dataset. This suggests that the choice of datasets can be decisive for the performance of neural networks — something that is already well recognized in the deep learning field — and that the lack of greater amounts of music available for training can lead to a surprising difference in performance. Therefore, it could prove useful to experiment with different datasets and hyperparameters in future works.

## 5. Conclusion

In this paper we performed the modeling, training and evaluation of a neural network model (B-RNN) which is capable of estimating the tempo (in bpm) of a musical piece. Providing a piece's Mel spectrogram as input, the output is a class as the estimation of tempo in bpm (beats per minute). We compared the results of the B-RNN with a CNN model considered state of the art [13], and with a replica of [13] which we retrained with the same datasets and parameters as the B-RNN for the purpose of evaluating the results (aiming to reduce bias). We proposed an analysis on the use of percussion-only dataset for training and the possibility of its use improving accuracy on tempo estimation.

Based on the results, it was noticed that the tempo estimation, overall, was more accurate for the percussion-only dataset. This suggests that tempo estimation can be more accurate over percussion-only tracks, but it is necessary to consider that only one dataset of this type was found available for the experiment, which was also used for both training and evaluation steps. Therefore, it is likely that the results could be biased in this direction. Further experiments should be made with more percussion datasets in order to gather stronger evidence. Still, the amount and variety of tracks in the Groove dataset (the only percussion dataset used) was considered good enough to be the first step on this kind of study.

The B-RNN model also presented very satisfactory results for the Accuracy2, which exceeded 85% of correct answers for almost all datasets. The values of Accuracy0 and Accuracy1, in general, for both B-RNN and CNNs, do not exceed 80%. Since these Accuracies are relevant to the understanding of music and its rhythm, it would be interesting to search for ways to refine this precision, so as to identify the correct multiple value of tempo. The B-RNN also had a better overall performance on the percussion-only dataset, which could suggest that recurrent networks identify temporal dependencies better than CNNs on percussion-only tracks. But further experiments should be done with more datasets for stronger evidence.

None of the models presented superior performance for absolutely every dataset, which suggests that the performance of a neural network in the estimation of tempo still depends a lot on the datasets that are selected for training, as well as the number of pieces used for training. Also, both original CNN model from [13] and its retrained replica presented quite different accuracies for the same dataset, which also enforces such suggestion, also possibly indicating that efforts in constructing more public dataset for research could be fruitful.

For future work, it would be interesting to experiment other hyperparameters for the network trainings; as well as to refine the B-RNN model, creating new layers for better signal processing, or even using B-LSTMs layers. It would also be interesting to create more datasets containing percussion lines-only, in order to reduce the possibility of bias in the assessment. Further experiments could be done to assess whether B-RNNs could be a better choice to identify temporal dependencies for percussion-only tracks.

**References**

[1] SCHEDL, M. *Automatically extracting, analyzing, and visualizing information on music artists from the World Wide Web.* PhD Thesis. Johannes Kepler University, Linz, 2008.

[2] GÓMEZ, E. et al. Music Information Retrieval: Overview, Recent Developments and Future Challenges. In: *17th International Society for Music Information Retrieval (ISMIR) Conference*, 2016, New York.

[3] ALONSO, M.; DAVID., B.; RICHARD, G. Tempo and beat estimation of musical signals. In: *5th International Society for Music Information Retrieval (ISMIR) Conference*, 2004, Barcelona.

[4] HÖRSCHLÄGER, F. et al. *Addressing Tempo Estimation Octave Errors in Electronic Music by Incorporating Style Information Extracted from Wikipedia*. Available on: <https://www.ifs.tuwien.ac.at/~knees/publications/hoerschlaeger_etal_smc_2015.pdf>. Accessed: 15 August 2020.

[5] HAINSWORTH, S.; MACLEOD, M.; Particle filtering applied to musical tempo tracking. *EURASIP J. on Applied Signal Processing*, v. 15, pp. 2385-2395, 2004.

[6] BÖCK, S. *Onset, Beat, and Tempo Detection with Artificial Neural Networks*. ca. 2010. Available on: <http://mir.minimoog.org/sb-diploma-thesis>. Accessed: 23 August 2019.

[7] BERRY, W. *Structural Functions in Music*. New Jersey: Prentice Hall, 1976. 447p.

[8] SCHREIRER, E.D. Tempo and beat analysis of acoustic musical signals. *The Journal of the Acoustical Society of America,* n. 103, v. 1, pp. 588-601, 1998.

[9] BÖCK, S.; KREBS, F.; WIDMER, G. Accurate Tempo Estimation based on Recurrent Neural Networks and Resonating Comb Filters. In: *16th International Society for Music Information Retrieval Conference*, Madrid, 2015.

[10] GKIOKAS, A.; KATSOUROS, V.; CARAYANNIS, G. Reducing tempo octave errors by periodicity vector coding and SVM learning. In: *13th International Society for Music Information Retrieval Conference*, 2012, Porto.

[11] HOCHREITER, S.; SCHMIDHUBER, J. Long short-term memory. *Neural computation*, v. 9, n. 8, pp. 1735-1780, 1997.

[12] BÖCK, S.; SCHEDL, M. Enhanced beat tracking with context-aware neural networks. *Proceedings of the 14th International Conference on Digital Audio Effects*, pp. 135- 139, September 2011.

[13] SCHREIBER, H.; MÜLLER, M. A Single-Step Approach to Musical Tempo Estimation Using a Convolutional Neural Network. In: *19th International Society for Music Information Retrieval Conference (ISMIR)*, 2018, Paris.

[14] BÖCK, S.; DAVIES, M.E.P. Deconstruct, analyse, reconstruct: How to improve tempo, beat, and downbeat estimation. *Proceedings of the 21st ISMIR Conference (International Society for Music Information Retrieval)*. Montreal, Canada, pp. 574-582, 2020.

[15] PEETERS, G.; FLOCON-CHLOET, J. Perceptual tempo estimation using GMM regression**.** *Proceedings of the second international ACM workshop on Music information retrieval with user-centered and multimodal strategies (MIRUM)***,** pp. 45-50, 2012.

[16] MARCHAND, U.; PEETERS, G. Scale and shift invariant time/frequency representation using auditory statistics: application to rhythm description. In: *IEEE International Workshop on Machine Learning for Signal Processing*, Salerno, September 2016.

[17] KNEES, P. et al. Two data sets for tempo estimation and key detection in electronic dance music annotated from user corrections. *Proceedings of the 16th International Society for Music Information Retrieval Conference (ISMIR)*, pp. 364-470, 2015.

[18] TZANETAKIS, G.; COOK, P. Musical genre classification of audio signals. *IEEE Transactions on Speech and Audio Processing*, v. 10, n. 5, pp. 293-302, 2002.

[19] RAFFEL, C. *Learning-Based Methods for Comparing Sequences, with Applications to Audio-to-MIDI Alignment and Matching*. PhD Thesis, Columbia University, New York, 2016.

[20] HOLZAPFEL, A. et al. Selective sampling for beat tracking evaluation. *IEEE Trans. on Audio, Speech, and Language Processing*, v. 20, n. 9, pp. 2539-2548, 2012.

[21] GOODFELLOW, I.; BENGIO, Y.; COURVILLE, A. *Deep learning*. Cambridge: MIT Press, 2016.

[22] MÜLLER, M. Fourier Analysis of Signals. In: MÜLLER, M. *Fundamentals of Music Processing*. Switzerland: Springer International Publishing, 2015, pp. 39-57.

[23] MONTECCHIO, N.; ROY, P.; PACHET, F. The Skipping Behavior of Users of Music Streaming Services and its Relation to Musical Structure. *arXiv:1903.06008*, 2019.

[24] SCHREIBER, H.; MÜLLER, M. Musical Tempo and Key Estimation using Convolutional Neural Networks with Directional Filters. *Proceedings of the 16th Sound and Music Computing Conference*, 2019.

[25] SCHUSTER, M.; PALIWAL K.K. Bidirectional Recurrent Neural Networks. *IEEE Transactions on Signal Processing*, v. 45, pp. 2673-2681, 1997.

[26] GRAVES, A.; MOHAMED, A.R.; HINTON, G. Speech recognition with deep recurrent neural networks. In: *Acoustics, speech and signal processing (ICASSP)*, 2013.

[27] HÖRSCHLÄGER, F. et al. *Addressing Tempo Estimation Octave Errors in Electronic Music by Incorporating Style Information Extracted from Wikipedia*. Available on: <https://www.ifs.tuwien.ac.at/~knees/publications/hoerschlaeger_etal_smc_2015.pdf>. Accessed: 15 August 2020.

[28] WU, F.H.F. (2015). Musical tempo octave error reducing based on the statistics of tempogram. In: *23rd Mediterranean Conference on Control and Automation (MED)*, 2015.